\documentclass  {aa}
\usepackage    {graphicx}
\usepackage    {txfonts}

\begin{document}


\title         {Studies of orbital parameters and pulse profile of the
                accreting millisecond pulsar XTE J1807-294}

\author        {M.G.F.\,Kirsch\inst{1}, K.\,Mukerjee\inst{4,5},
                M.G.\,Breitfellner\inst{1}, S.\,Djavidnia\inst{1},
                M.J.\,Freyberg\inst{2}, E.\,Kendziorra\inst{3}
                and M.J.S.\,Smith\inst{1} }

\offprints     {M. Kirsch mkirsch@xmm.vilspa.esa.es}

\institute     {European Space Agency (ESA), Research and Scientific Support
                Department (RSSD), Science Operations and Data System Division
                (SCI-SD), {\em XMM-Newton\/} Science Operations Centre, Apartado - 
                P.O. Box 50727, E-28080 Madrid, Spain
                \and
                Max-Planck-Institut f\"{u}r extraterrestrische Physik,
                Giessenbachstrasse, D-85748 Garching, Germany
                \and
                Institut f\"{u}r Astronomie und Astrophysik Universit\"{a}t
                T\"{u}bingen, Abteilung Astronomie, Sand 1, D-72076
                T\"{u}bingen, Germany
                \and
                Space Research Centre, Department of Physics and Astronomy,
                University of Leicester, Leicester, LE1 7RH, UK
                \and
                Department of Astronomy and Astrophysics, Tata Institute of
                Fundamental Research, Colaba, Mumbai-400005, India }

\date          {Received xx February 2004; accepted xx xxxxx 2004}

\abstract      {
The accreting millisecond pulsar XTE~J1807-294 was observed as 
a Target of Opportunity (ToO) by {\em XMM-Newton\/} on March 22, 2003 after its
discovery on February 21, 2003 by {\em RXTE\/}.
The source was detected in its bright phase with an observed average count
rate of 33.3\,cts s$^{-1}$ in the {\em EPIC-pn\/} camera in the 0.5--10\,keV energy
band (3.7\,mCrab). Using the earlier established best-fit
orbital period of 40.0741$\pm$0.0005 minutes from the {\em RXTE\/} observations and 
considering a circular binary orbit as first approximation, we derived a
value of 4.8$\pm$0.1\,lt-ms for the projected orbital radius of the binary system and 
an epoch of the orbital phase of MJD\,52720.67415(16). The barycentric 
mean spin period of the pulsar was derived as 5.2459427$\pm$0.0000004\,ms. 
The pulsar's spin-pulse profile showed a prominent (1.5\,ms FWHM) 
pulse, with energy and orbital phase
dependence in the amplitude and shape.
The measured pulsed fraction in four energy bands was found to be 3.1$\pm$0.2\,\% (0.5--3.0\,keV), 
5.4$\pm$0.4\,\% (3.0--6.0\,keV),
5.1$\pm$0.7\,\% (6.0--10.0\,keV) and 3.7$\pm$0.2\,\% (0.5--10.0\,keV), 
respectively.
Studies of spin-profiles with orbital phase and energy showed significant
increase in its pulsed fraction during the second observed orbit of the
neutron star, gradually declining in the subsequent two orbits, which was
associated with sudden but marginal increase in mass accretion.
From our investigations of orbital parameters and
estimation of other properties of this compact binary system, 
we conclude that XTE~J1807-294 is very likely a candidate for a millisecond
radio pulsar.
               }

\authorrunning {M. Kirsch et al.}
\titlerunning  {XTE~J1807-294}
\maketitle


\section       {Introduction}
\label         {sec:intro}


The accreting millisecond pulsar XTE~J1807-294 was discovered as the fourth 
candidate of this class by {\em RXTE\/}
(Markwardt, Smith \& Swank \cite{markwardt_8080}) on February 21, 2003.
The coherent pulsation of 5.245902\,ms was detected and subsequently 
the  orbital period of 40.0741$\pm$0.0005 minutes was established 
(Markward, Juda \& Swank \cite{markwardt_8095}), confirming it to be the
shortest orbital period out of five such pulsar systems discovered so far.
The spectral measurements of XTE~J1807-294 showed a combination of absorbed
black-body and a comptonization model with no absorption or emission lines
(Campana et al. \cite{campana2003}). 
The other accreting millisecond pulsars known so far are
SAX\,J1808.4-3658 (2.49\,ms; Wijnands \& van der Klis \cite{wijnands1998}),
XTE~J1751-305 (2.30\,ms; Markwardt \& Swank \cite{markwardt_7867}),
XTE~J0929-314 (5.41\,ms; Remillard, Swank \& Strohmayer \cite{remillard2002}),
XTE~J1807-294 (5.25\,ms; Markwardt, Smith \& Swank \cite{markwardt_8080})
and the most recent XTE~J1814-338
(3.18\,ms; Strohmayer et al. \cite{strohmayer2003}).
Detailed studies on the first three candidates showed that these were short
period, X-ray transient, ultra compact binary systems having orbital periods
$\leq$ 80 minutes (Bildsten \& Chakrabarty \cite{bildsten2001};
Juett, Galloway \& Chakrabarty \cite{juett2003}; 
Markwardt et al. \cite{markwardt_apj}).
The binary parameters imply $\approx $0.01--0.02\,$M_{\odot}$ 
white dwarf donors with moderately high inclination in these binaries 
(Galloway et al. \cite{galloway2002}; Bildsten \cite{bildsten2002};
Markwardt et al. \cite{markwardt_apj}).


Studies of the binary parameters of compact stars and their stellar companions
provide a better understanding of the physical conditions of this new class
of objects, their accretion physics and possible evolutionary connections to
the class of radio pulsars
(Bhattacharya \& van den Heuvel \cite{bhattacharya1991}).
With this in mind, we have done a detailed analysis of the ToO data
of XTE~J1807-294 obtained by {\em XMM-Newton\/} on March 22, 2003. The spin-period
of XTE~J1807-294 has been confirmed, using {\em XMM-Newton\/} data and the 
projected semi-major axis of the binary orbit derived using a technique 
applicable to short period pulsars (Kirsch \& Kendziorra \cite{kirsch2003}).
We present new results on XTE~J1807-294 in this letter from our studies of
orbital parameters, orbital phase and energy dependent pulse profiles,
which provide better insight into the physical conditions of this pulsar system.

\section       {XMM-Newton Observations}
\label         {sec:obs}

XTE~J1807-294 was observed by {\em XMM-Newton\/} on March 22, 2003, starting at
2003-03-22 13:40:27\,UT, with Obs-Id 01579601 in revolution number 601.
The {\em EPIC-pn\/} CCD camera was operated in {\em Timing\/} 
mode with the {\em Thick Filter\/}.
For this analysis, data from a 9\,293\,s long 
exposure were used which covered almost four orbital periods of the source. 

The {\em EPIC-pn\/} camera provides high time resolution 
in {\em Timing\/} mode (0.03\,ms), and {\em Burst\/} mode (7\,$\mu$s) with moderate energy 
resolution (E/dE = 10--50) in the 0.1--15\,keV energy band. This makes 
the pn-camera most suitable for simultaneous timing and spectral studies of millisecond 
pulsar sources.

\section       {Data analysis}
\label         {sec:analysis}

\subsection    {Data reduction}
\label         {subsec:reduce}

The data were processed with \textit{SAS 5.4.1}. Event times 
were corrected to the solar system barycentre with the {\em SAS\/} tool \textit{barycen}. 
In the {\em Timing\/} mode of the pn camera, a point source will be completely smeared 
out in the DETY
direction. Hence, this mode provides only one-dimensional spatial information.
Therefore, we used a nine column (37\, arcsec) wide source 
extraction region centred on the source, consisting of CCD 
columns 33--41. CCD columns 03--11 were used for the 
background extraction region.
An average counting rate of 33.7\,cts s$^{-1}$ was detected from the source 
in the 0.5--10\,keV band (3.7\,mCrab), while an average background of
0.4\,cts s$^{-1}$ was detected.

\subsection    {$\chi^{2}$ grid search of the orbital parameters} 
\label         {subsec:chi}

Using the best fit orbital period of 40.0741$\pm$0.0005 minutes
(Markwardt, Juda \& Swank \cite{markwardt_8095}),
and assuming a circular orbit, the relevant orbital parameters
were determined through maximum $\chi^{2}$ epoch folding using 
6 bin pulse profiles.
\begin         {figure} [hbt]
\resizebox     {\hsize}{!}{\includegraphics {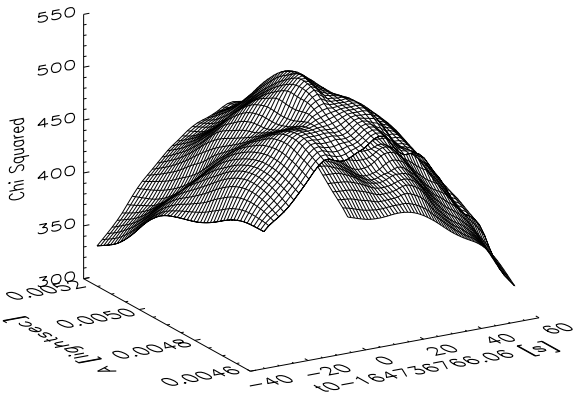}}
\caption       {
$\chi^{2}$ grid search of the orbital parameters. See text for details. 
               }
\label         {fig:chi_tri}
\end           {figure}
To this end, the photon arrival times with respect to the binary barycentre
of all events were corrected for different
values of projected orbital radius $x_{0}$, and orbital phase $t_{0}$,
where $x_{0}$ was varied in seven steps of 0.10\,light-ms and $t_{0}$ in 14 steps equivalent
to 1.0$^{\circ}$, using
$t_{\mbox{\scriptsize binary corrected}}
       =t-x_{0}\sin(2\pi/P_{\mbox{\scriptsize orb}} \times (t -t_{0}))
$.

The initial parameter values were obtained by grouping events into 20
phase bins of the binary orbit and identifying two phases separated by 180$^{\circ}$ via a
$\chi^{2}$ epoch folding for each of the phase bins.

The results of the two-dimensional parameter space search are shown in 
Figure~\ref{fig:chi_tri}.
The data were interpolated through Delaunay triangulation in order to better to visualise the
$\chi^{2}$ dependence on $x_{0}$ and $t_{0}$.
A maximum $\chi^{2}$ of 520 was found for $x_{0}$ of
4.8 $\pm$0.1\,light-ms and $t_{0}$ of 52720.67415(16) (MJD).

In order to test the assumption of a circular orbit, an analysis of possible 
orbital eccentricity of the neutron star orbit was carried out. However,
no statistical significance for an eccentric orbit could be found
and therefore the orbit is treated as circular in all further analysis.
Table 1 summarizes relevant parameters of the accreting
millisecond pulsar XTE~J1807-294 established so far.

\begin         {table} [h]
  \begin       {footnotesize}
    \begin     {center}
    \caption[] {Measured parameters of XTE~J1807-294.}
      \begin   {tabular}{lcc}
      \hline
      \noalign {\smallskip}
      Parameter            & Value                   & Instrument \\
      \hline
      \noalign {\smallskip}
      \hline
      \noalign {\smallskip}
      RA                   &  18:07:00.0 (J2000)   & {\em Chandra\/}$^{1}$ \\
      DEC                  & -29:24:30.0 (J2000)     & {\em Chandra\/}$^{1}$ \\
      Orbital  period        & 40.0741(5) min            & {\em RXTE\/}$^{1}$ \\
      Spin period          & 5.2459427(4) ms         & {\em XMM-Newton\/}$^{2}$ \\
      Proj. orbital radius $x_0$ & 4.8(1) light-ms         & {\em XMM-Newton\/}$^{2}$ \\
      Orbital phase $t_0$      & 52720.67415(16) (MJD)         & {\em XMM-Newton\/}$^{2}$\\
      \noalign {\smallskip}
      \hline
      \end     {tabular}
    \end       {center}
    $^{1}$Markward, Juda \& Swank \cite{markwardt_8095},
    $^{2}$this work

  \end         {footnotesize}
\label         {tab:xtej1807}
\end           {table}

\subsection    {Pulse profile}
\label         {subsec:pulse}

\begin         {figure} [t]
\resizebox     {\hsize}{!}{\includegraphics [angle=0]{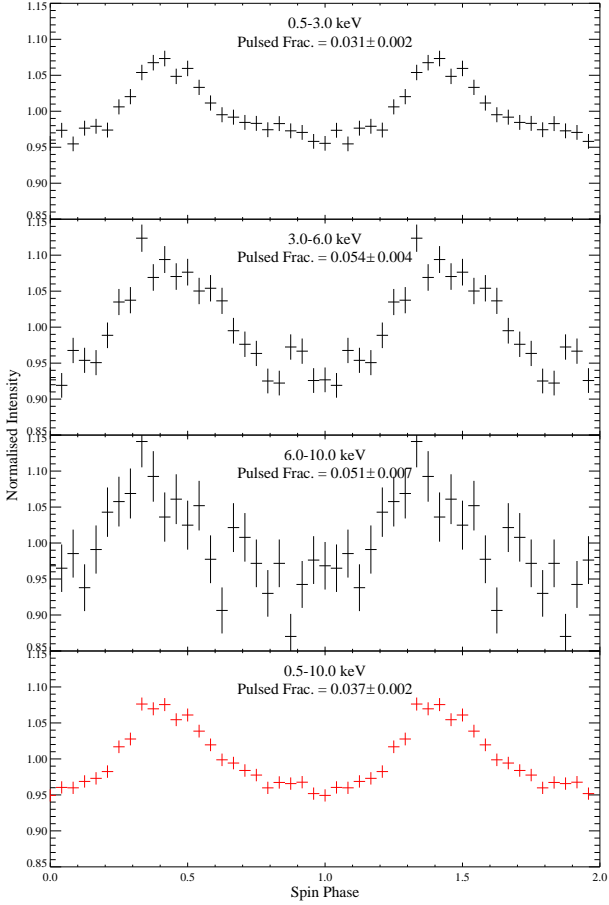}}
\caption       {Spin-pulse profiles of XTE~J1807-289 
		in different energy bands.} 
\label         {fig:flc1}
\end           {figure}

Using the best-fit orbital period of 40.0741$\pm$0.0005 minutes 
(Markwardt, Juda \& Swank \cite{markwardt_8095}) and the 
orbital parameters derived as described in Section~\ref{subsec:chi},
photon arrival times were corrected to the binary barycentre. We 
then derived the barycentric mean spin period of the pulsar of
5.2459427$\pm$0.0000004\,ms.
By epoch folding the data with the derived period with respect to the epoch at 
MJD~52720.72457, we established spin pulse profiles in four different energy bands
between 0.5--10\,keV as shown in Figure~\ref{fig:flc1}. These profiles show
a very prominent pulse (1.5\,ms FWHM).
The shapes and relative strengths of the profile varies with energy.
We determined the pulsed fraction in these energy bands using the expression

$$
\mbox{pulsed fraction}
	= \frac{ \sum_{\mbox{\scriptsize bins}} \left|\mbox{flux per bin $-$ mean flux per bin}\right|}{\mbox{total flux}}
$$

The measured pulsed fraction was found to be 3.1$\pm$0.2\,\% (0.5--3.0\,keV), 
5.4$\pm$0.4\,\% (3.0--6.0\,keV),
5.1$\pm$0.7\,\% (6.0--10.0\,keV) and 3.7$\pm$0.2\,\% (0.5--10.0\,keV), 
respectively, in the four different energy bands as in
Figure~\ref{fig:flc1}. 

\begin         {figure} [t]
\resizebox     {\hsize}{!}{\includegraphics [angle=0]{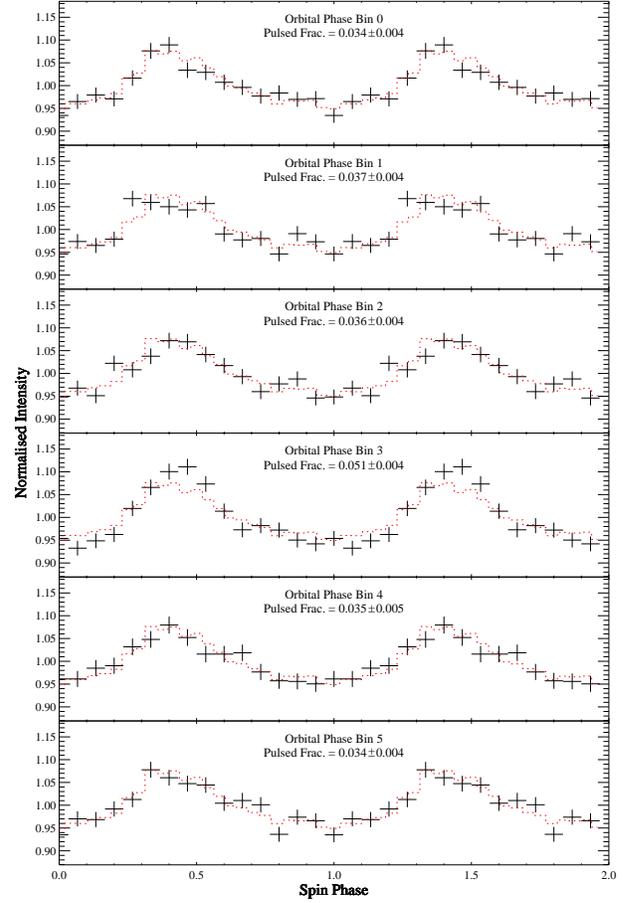}}
\caption       {Spin-pulse profiles of XTE~J1807-289 at different orbit phases
                in the energy range 0.5--10\,keV. In dotted red the orbit averaged folded
		spin light curve from Figure~\ref{fig:flc1} in the same energy range.} 
\label         {fig:flc2}
\end           {figure}

\begin         {figure} [t]
\resizebox     {\hsize}{!}{\includegraphics [angle=0]{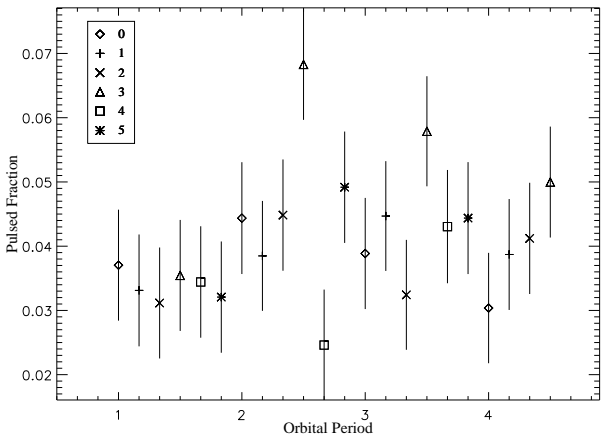}}
\caption       {Pulsed fraction values for 6 orbital phases indicated by different 
		symbols for all the four orbits in the 0.5--3.0\,keV energy band.}
\label         {fig:folding}
\end           {figure}

To study the variation of the spin-pulse profile with orbital 
phase, we grouped the data in six different phases covering
the complete binary orbit, where phase 0 starts at
$t_0-44.08\,s$. The folded light curves corresponding to six
orbital phases derived for the 0.5--10.0\,keV energy band are shown in 
Figure~\ref{fig:flc2}. The spin-pulse profile corresponding to orbital phase~3,
shows a significantly higher pulsed fraction. 
We then performed a detailed analysis for each of the four observed orbits
and detected a significantly higher pulsed fraction corresponding to orbital
phase~3 of orbit 2, which gradually declines during orbits 3 \& 4.
Figure~\ref{fig:folding} gives the pulsed fraction calculated for six 
orbital phases for all the four binary orbits in the 0.5--3.0\,keV energy band.

\section       {Discussion}
\label         {sec:discussion}

X-ray pulse profiles of XTE~J1807-294 showed both energy and orbital phase
dependence during the {\em XMM-Newton\/} observation. 
Pulse profiles of XTE~J1807-294 show orbital phase dependent changes
and variation of pulsed fraction as quoted in Figure~\ref{fig:flc2}.
These effects are also seen in
other X-ray pulsars due to geometrical effects with respect to the line of
sight and physical conditions of the pulsar
(Wang \& Welter \cite{wang1981}).
The sudden increase of the pulsed fraction value, as seen in Figure~\ref{fig:folding},
particularly during phase~3 in orbit~2, decreases gradually
during subsequent orbits and is caused very likely by a
sudden but marginal increase in mass accretion. This causes an increase
in the pulsed fraction of the pulsar which then gradually decreases due to the
reduction of accreting matter and hence the pulsed fraction when neutron star 
poles are visible at the line of sight corresponding to a particular orbit
phase~3.
Such changes in the pulse profiles and pulsed fraction due to increase in
mass accretion rate during an outburst phase are commonly seen in X-ray
binaries (Galloway et al. \cite{galloway2002};
Mukerjee et al. \cite{mukerjee2000}).

Determination of binary orbital parameters enables us to estimate properties of 
the accreting binary system. The short orbital period of 40.0741 minutes 
of XTE~J1807-294 immediately confirms that it is the most compact binary 
system out of the five accreting millisecond pulsars known so far. Estimations 
of the semi-major axis $a \sin i$ for this binary using our technique
described above give 0.0048 lt-s which enables us to estimate the
mass function of this pulsar using the expression,

$
f_{M} = 1.074 \times 10^{-3} P^{2/3} {(a \sin i / 1\,\mbox{lt-s})}^{3} 
{(P_{\mbox{\scriptsize orb}} / 1\,\mbox{d})}^{-2}M_{\odot}
$.

The pulsar mass function $f_{M}=1.6 \times 10^{-7} M_{\odot}$ is the smallest
presently known for any stellar binary. For a neutron star mass of
$1.4\,M_{\odot}$, the minimum companion mass would be $0.007\,M_{\odot}$
for an inclination angle of $i=90^{\mbox{\scriptsize o}}$,
suggesting that the companion of XTE J1807-294 is in the regime of very 
low mass dwarf, of the order of 7 Jupiter masses. The nature of the companion
 in an ultracompact binary system depends on prior evolution scenario. 
Binary systems with $P_{orb}$ $\leq$ 80 minutes can be formed in at least two possible 
evolution as summarized by Deloye \& Bildsten \cite{del2003} 
and the references therein. The first is, stable mass transfer 
on to a neutron star from an evolved main-sequence star or a He-burning star. 
In this scenario, the main sequence star, when its core has nearly completed 
Hydrogen burning, is brought into Roche lobe contact due to loss of orbital 
angular momentum from magnetic braking which can evolve into a system with orbital 
periods comparable to those of accreting millisecond pulsars detected so far and 
can reach a minimum  period of 10 minutes 
(Nelson, Rappaort \&  Joss \cite{nelson1986}, Podsiadlowski et al. \cite{podsi2002}; Nelson \& Rappaport \cite{nelson2003}) and  when the system evolving 
through 40 minutes on the way out from the minimum period can have masses (0.01 $M_{\odot}$) more 
in line with the measurements (Nelson \& Rappaport \cite{nelson2003}). The second channel involves 
evolution through common-envelope phase during an unstable mass transfer from the 
white dwarf donor's progenitor star of either pure He or C/O  and are most recently  discussed 
in detail by Deloye \& Bildsten (\cite{del2003}) . 

The composition of the donor in XTE J1807-294 could be He/H, pure He 
or a C/O mixture depending on formation history. Without further data, the composition 
of the donor can not be determined. However, one can examine the nature of the donor based 
on orbital information of the binary system by applying models developed for ultra compact 
accreting millisecond pulsars by Deloye \& Bildsten (\cite{del2003}). 
Their model considers low-mass $\le$ 0.1 $M_{\odot}$ White dwarf companion of arbitrary degeneracy 
of evolved He or C/O composition and central temperature of $10^{5}$--$10^{7}$ as range of 
relevance for these objects, the corresponding central densities ($10^{3} g cm^{-3}$ are such that 
Coulomb and thermal contributions to the equation of state provide non-negligible corrections 
to degenerate electron pressure, affecting their M-R relations. The model thus based on equation 
of state, adequately describe relevant physics and yields analytic description of the qualitative 
behavior of M-R relations and how they  are affected by Coulomb and thermal contributions 
to the equation of state. The model established a relationship between the orbital 
inclination for ultra compact binaries and the donor's composition and central core 
temperature. The application of this model to three of the known accreting millisecond pulsar 
systems suggest that if the donors in all the three systems are He White dwarfs, the T=0 objects 
are allowed in XTE J0929-314 and XTE J1807-294 while XTE J1751-305 requires a hot donor 
(Bildsten \cite{bildsten2002}). While C/O cold donor is possible for XTE J1807-294, the other two systems  
required hot C/O donors. For XTE J1807-294 in particular, with measured orbital parameters 
suggest that the companion for this ultra compact binary system has greater phase space 
available for a C/O donor than a He donor. Because for XTE J1807-294, the apriori probability 
greatly favors C/O donor in terms of allowed inclination as the probability that a donor 
is He is only 15 \% in this system (Deloye \& Bildsten \cite{del2003}). Further evolution of this system 
will depend on its orbital inclination which in terms depends on donor mass, core temperature 
and its composition. From known properties of this binary system, it is likely that at the 
end of its accretion phase, it could be a candidate for a millisecond radio pulsar.

\begin         {acknowledgements}

The XMM-Newton project is an ESA Science Mission with instruments
and contributions directly funded by ESA Member States and the
USA (NASA). The German contribution of the XMM-Newton project is
supported by the Bundesministerium f\"{u}r Bildung und
Forschung/Deutsches Zentrum f\"{u}r Luft- und Raumfahrt.

We thank an anonymous referee for his very constructive comments.
\end           {acknowledgements}

\end           {document}